\documentclass{article}
\usepackage{spconf,amsmath,graphicx}

\usepackage{amsmath,amsfonts,bm}

\def\eqref#1{equation~\ref{#1}}

\def\1{\bm{1}}

\DeclareMathAlphabet{\mathsfit}{\encodingdefault}{\sfdefault}{m}{sl}
\SetMathAlphabet{\mathsfit}{bold}{\encodingdefault}{\sfdefault}{bx}{n}

\usepackage{hyperref}
\usepackage{url}

\usepackage{xcolor}
\newcommand{\KGnote}[1]{{\color{blue}{\bf KG: }#1}}

\newcommand{\KG}[1]{{\color{black}{}#1}}
\newcommand{\KGcr}[1]{{\color{blue}{}#1}}

\newcommand{\KGiclr}[1]{{\color{black}{}#1}}
\newcommand{\KGs}[1]{{\color{blue}{}#1}}

\newcommand{\CA}[1]{{\color{black}{}#1}}
\newcommand{\CAn}[1]{{\color{brown}{}#1}}
\newcommand{\CAr}[1]{{\color{cyan}{}#1}}
\newcommand{\CAcut}[1]{{\color{green}{}#1}}
\newcommand{\CAis}[1]{{\color{cyan}{}#1}}
\newcommand{\ISnew}[1]{{\color{brown}{}#1}}

\renewcommand{\KGiclr}[1]{{\color{black}{}#1}}

\renewcommand{\CAn}[1]{{\color{black}{}#1}}
\renewcommand{\CAr}[1]{{\color{black}{}#1}}
\renewcommand{\CAis}[1]{{\color{black}{}#1}}
\renewcommand{\KGs}[1]{{\color{black}{}#1}}
\renewcommand{\ISnew}[1]{{\color{black}{}#1}}
\renewcommand{\KGcr}[1]{{\color{black}{}#1}}

\renewcommand{\CAcut}[1]{{}}

\renewcommand{\KGnote}[1]{{\color{blue}{}}} 

\usepackage[utf8]{inputenc} 
\usepackage[T1]{fontenc}    
\usepackage{hyperref}       
\usepackage{url}            
\usepackage{booktabs}       
\usepackage{amsfonts}       
\usepackage{nicefrac}       
\usepackage{microtype}      
\usepackage{enumitem}
\usepackage{graphicx}
\usepackage{siunitx}
\bfseries
\usepackage{soul}
\usepackage{subcaption}
\usepackage[nodisplayskipstretch]{setspace} 
\title{Learning Audio-Visual Dereverberation}

\name{Changan Chen$^{1,2}$ \hspace{3mm} Wei Sun$^{1}$ \hspace{3mm} David Harwath$^{1}$ \hspace{3mm} Kristen Grauman$^{1,2}$
}
\address{$^1$UT Austin \hspace{3mm} $^2$FAIR, Meta AI}

\begin{document}
\ninept

\maketitle

\begin{abstract}
Reverberation\CAcut{from audio reflecting off surfaces and objects in the environment} not only degrades the quality of speech for human perception, but also severely impacts the accuracy of automatic speech recognition. Prior work attempts to remove reverberation based on the audio modality only. Our idea is to learn to dereverberate speech from audio-visual observations. The visual environment surrounding a human speaker reveals important cues about the room geometry, materials, and speaker location, all of which influence the precise reverberation effects. 
We introduce Visually-Informed Dereverberation of Audio (VIDA), an end-to-end approach that learns to remove reverberation based on both the observed monaural sound and visual scene.  In support of this new task, we develop a large-scale dataset \CAis{SoundSpaces-Speech} that uses realistic acoustic renderings of speech in real-world 3D scans of homes offering a variety of room acoustics.  Demonstrating our approach on both simulated and real imagery for speech enhancement, speech recognition, and speaker identification, we show it achieves state-of-the-art performance and substantially improves over audio-only methods. 
\end{abstract}

\section{Introduction}

Audio reverberation occurs when multiple reflections from surfaces and objects in the environment build up then decay, altering the original audio signal.  While reverberation bestows a realistic sense of spatial context, it also can degrade a listener's experience.  In particular, the quality of human speech is greatly affected by reverberant environments---as illustrated by how difficult it can be to parse the words of a family member speaking loudly from another room in the house, a tour guide describing the artwork down the hall of a magnificent cavernous cathedral, or a colleague participating in a Zoom call from a cafe.  Consistent with the human perceptual experience, automatic speech recognition (ASR) systems noticeably suffer when given reverberant speech input~\cite{reverb-challenge,zhao_monaural_2020,szoke_2019,han_learning_2015,wu_end--end_2017,ernst_speech_2019}.  
Thus there is great need for intelligent \emph{dereverberation} algorithms that can strip away reverb effects for speech enhancement, recognition, and other downstream tasks, which could in turn benefit many applications in teleconferencing, assistive hearing devices, augmented reality, and video indexing.

The audio community has made steady progress devising machine learning solutions to tackle speech dereverberation~\cite{ernst_speech_2019,giri_improving_2015,xvector,han_learning_2015,wu2016,zhao2019,su_hifi-gan_2020}.
The general approach is to take a reverberant speech signal, usually represented with a Short-Time Fourier Transform (STFT) spectrogram, and feed it as input to a model that estimates a clean version of the signal with the reverberation removed. Past approaches have tackled this problem with signal processing and statistical techniques~\cite{wpe,naylor_gaubitch_2010}, while many modern approaches are based on neural networks that learn a mapping from reverberant to clean spectrograms~\cite{han_learning_2015,ernst_speech_2019,fu_metricgan_2019}. To our knowledge, all existing models for dereverberation rely purely on audio. 
Unfortunately this often underconstrains the dereverberation task since the latent parameters of the recording space are not discernible from the audio alone.

\begin{figure}[t]
    \centering
    \includegraphics[width=\linewidth]{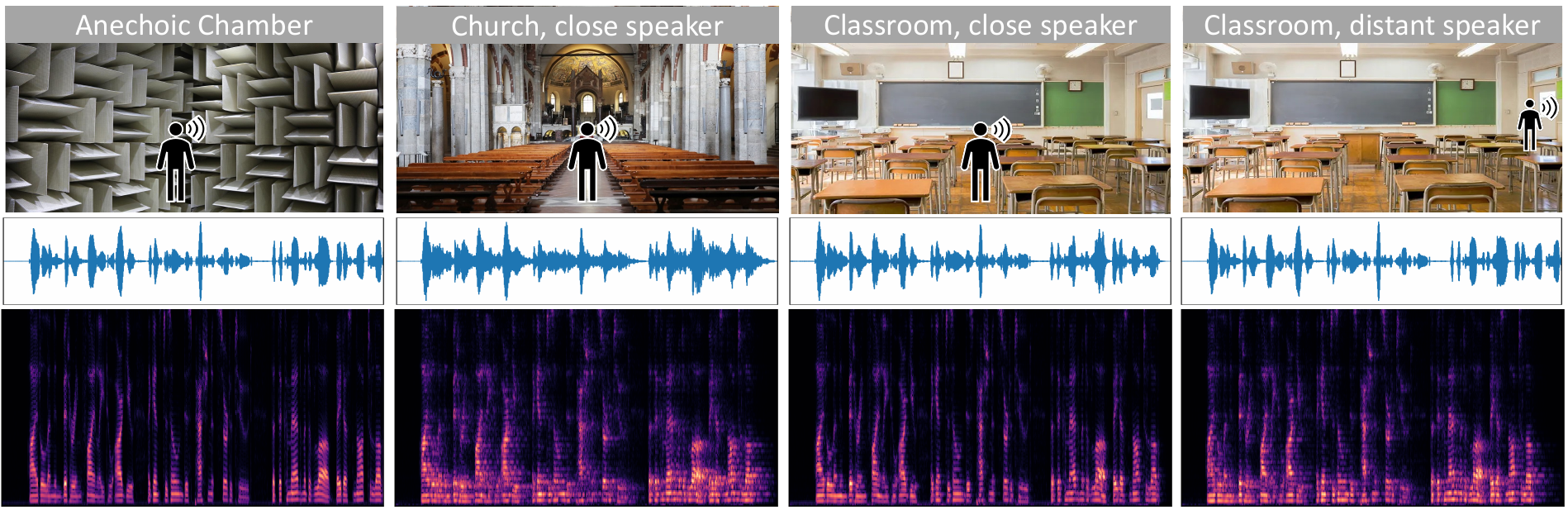}
    \caption{Visual cues reveal key factors influencing reverb effects on human speech audio.  For example, these audio speech samples (depicted as \CA{waveforms and} spectrograms) are identical lexically, but have very different reverberation properties owing to their differing environments.  In the church, reverb is strong, in the classroom it is less, and when the speaker is distant from the camera it is again more evident. 
    }
    \vspace*{-0.2in}
    \label{fig:visual_cues}
\end{figure}

However, we observe that in many practical settings of interest---video conferencing, augmented reality, Web video indexing---reverberant audio is naturally accompanied by a visual (video) stream.  Importantly, the visual stream offers valuable cues about the 
room acoustics affecting reverberation: where are the walls, how are they shaped, where is the human speaker, what is the layout of major furniture,  what are the room's dominant materials (which affect absorption), and even what is the facial appearance and/or body shape of the person speaking (since body shape determines the acoustic properties of a person's speech, and reverb time is frequency dependent). For example, reverb is typically stronger when the speaker is further away; speech is more reverberant in a large church or hallway; heavy carpet absorbs more sound.  See Figure~\ref{fig:visual_cues}.
While some recent works explore acoustic modeling using images~\cite{majumder2022fewshot,chen22vam,luo2022learning,kim2020acoustic}, no prior work has investigated how to leverage visual-acoustic cues for dereverberation.

\begin{figure}[t]
    \centering
    \includegraphics[width=0.99\linewidth]{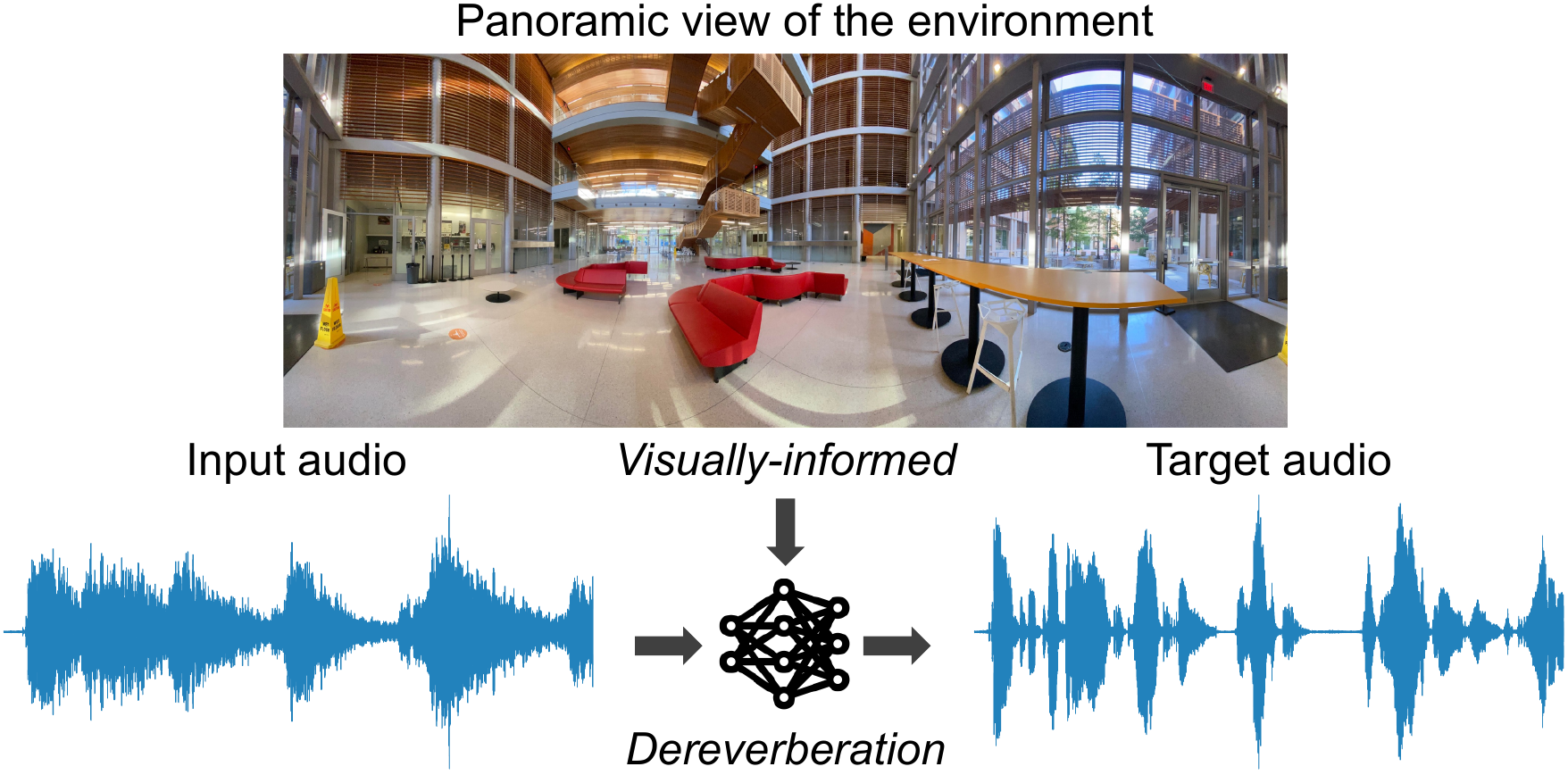}
    \vspace*{-0.1in}
    \caption{The goal of audio-visual dereverberation is to leverage the visual observation of the environment to improve speech enhancement.}
    \vspace*{-0.2in}
    \label{fig:concept}
\end{figure}

Our idea is to learn to dereverberate speech from audio-visual observations (Fig.~\ref{fig:concept})  In this task, the input is reverberant speech and visual observations of the environment surrounding the human speaker, and the output is a prediction of the clean source audio.  To tackle this problem, there are two key technical challenges.  First, how to model the multi-modal dereverberation process in order to infer the latent clean audio.  Second, how to secure appropriate training data spanning a variety of physical environments for which we can sample speech with known ground truth (non-reverberant, anechoic) audio. The latter is \KG{also} non-trivial because ordinary audio/video recordings are themselves corrupted by reverberation but lack the ground truth source signal we wish to recover. 

For the modeling challenge, we introduce an end-to-end approach called 
Visually-Informed Dereverberation of Audio (VIDA). 
VIDA consists of a Visual Acoustics Network (VAN)
that learns reverberation properties of the room geometry, object locations, and speaker position. Coupled with a  multi-modal UNet dereverberation module, it learns to remove the reverberations from a single-channel audio stream. In addition, we propose an audio-visual (AV) matching loss to enforce consistency between the visually-inferred reverberation features and those inferred from the audio signal.
We leverage the outputs of our model for multiple downstream tasks: speech enhancement, speech recognition, and speaker identification.
 
Next, to address the training data challenge, 
we develop SoundSpaces-Speech, a new large-scale audio-visual dataset based on SoundSpaces~\cite{chen_soundspaces_2020}, a 3D simulator for real-world scanned environments that allows both visual and acoustic rendering.  
Our data approach inserts ``clean" audio voices together with a 3D humanoid model at various positions within an array of indoor environments, then samples the images and properly reverberating audio when placing the receiver microphone and camera at other positions in the same house.  This strategy allows sampling realistic audio-visual instances coupled with ground truth raw audio to train our model, and it has the added benefit of allowing controlled studies that vary the parameters of the capture setting.    \KG{As we will show, the data also supports sim2real transfer for applying our model to real audio-visual observations.}  

Our main contributions are to 1) present the task of audio-visual dereverberation, 2) address it with a new multi-modal modeling approach and a novel reverb-visual matching loss, 3) provide a benchmark evaluation framework built on both SoundSpaces-Speech and real data, and 4) demonstrate the utility of AV dereverberation for multiple practical tasks.
We first train and evaluate our model on 82 large-scale real-world environments---each a multi-room home containing a variety of objects---coupled with speech samples from the LibriSpeech dataset~\cite{librispeech}.
We consider both near-field and far-field settings where the human speaker is in-view or quite far from the camera, respectively.  The proposed model outperforms methods restricted to the audio stream, and improves the state of the art for \KG{multiple tasks with} speech enhancement. We also 
show that our model trained in simulation can transfer directly to real-world data.
Overall, our work shows the potential for speech enhancement models to benefit from seeing the 3D environment.

\section{Related Work}

\noindent\textbf{Audio dereverberation and speech enhancement.} 
Audio dereverberation and speech enhancement have a long and rich literature~\cite{neely_allen_1979,miyoshi_kaneda_1988,naylor_gaubitch_2010,reverb-challenge,benetsy_speech_enhancement}. While 
dereverberation can be done with microphone arrays, we focus on single audio channel approaches, \KG{which require fewer assumptions about the input data}. Recent deep learning methods achieve promising results to dereverberate~\cite{han_learning_2015,wu2016,zhao2019,ernst_speech_2019,zhao_monaural_2020,su_hifi-gan_2020}, denoise~\cite{vondrick-denoising-speech-nips2019,fu_metricgan_2019,su_hifi-gan_2020}, or separate~\cite{hershey2016deep,stoller2018adversarial} the audio stream using audio input alone,  
\KG{and such enhancements can improve downstream speech recognition~\cite{ko2017study,reverb-challenge} and speaker recognition~\cite{xvector}.} 
Acoustic simulations can provide data augmentation during training~\cite{reverb-challenge,han_learning_2015,ko2017study,zhao_monaural_2020}. 
Accounting for environmental effects on reverb, some work targets ``room-aware" deep audio features capturing reverberation properties (e.g., RT60)~\cite{giri_improving_2015}, or injects reverberation effects from a different room via acoustic matching~\cite{su2020acoustic}.  
To our knowledge, the only prior work drawing on the \emph{visual} stream to infer dereverberated audio is limited to using lip regions on near-field faces to first separate out distractor sounds~\cite{tan_audio-visual_2020}, and does not model anything about the visual scene for dereverberation purposes.  In contrast, our model accounts for the full visual scene, far-field speech sources, and even out-of-view speakers. \KG{Our approach is the first to learn visual room acoustics for dereverberation, and it yields state-of-the-art  results with direct benefits for multiple downstream tasks.}

\noindent\textbf{Visual-acoustic learning.} 
The room impulse response (RIR) is the transfer function capturing the room acoustics for arbitrary source stimuli; once convolved with a sound waveform, it produces the sound of that source in the context of the particular physical space. While traditionally measured 
with specialized equipment in the room itself~\cite{stan2002comparison,Holters_rir} or else simulated with  sound propagation models \CA{(e.g., geometric~\cite{image_method,chen_soundspaces_2020} or wave-based ~\cite{Murphy_waveguide})},
recent work explores estimating an RIR from an input image \KG{using} CNNs~\cite{kon_estimation_2019} or conditional GANs~\cite{singh_image2reverb_2021} in order to simulate reverberant sound for a given environment. Video-based methods have also explored ways to lift monaural audio into its spatialized (binaural, ambisonic) counterpart in order to create an immersive audio experience for human listeners~\cite{morgado-2018,25d-visual-sound,scene-aware-360}.
Such methods share our interest in learning visual properties of a scene that influence the audio channel.  However, unlike any of the above methods, rather than generate spatialized audio to benefit human listeners in augmented or virtual reality, our goal is to dereverberate audio---removing the effects of the room acoustics---
\KG{to benefit}
automatic speech analysis.
In addition, prior methods use imagery taken at camera positions at an unknown offset from the microphone, i.e., \KG{conflating all RIRs for a scene with one image}, which limits 
them to a coarse characterization of the environment~\cite{jeub_binaural,jeub_air,open_air}.  In contrast,  our data and model align the camera and microphone to capture novel fine-grained audio-visual properties, including the human speaker's location with respect to the microphone \KG{when the speaker is in view}.

\noindent\textbf{Audio-visual simulations.}
Recent work in embodied AI explores how vision and sound together can help agents move intelligently in 3D environments.  Driven in part by new tools for audio-visual (AV) simulations in realistic scanned environments~\cite{chen_soundspaces_2020,chen22soundspaces2}, new research develops deep reinforcement learning approaches to train agents to navigate to sounding objects~\cite{chen_soundspaces_2020,gan2019look,chen_savi_2021,chen_waypoints_2020}, explore unmapped environments~\cite{dean-curious-nips2020}, or move around to better separate multiple overlapping sounds in a house~\cite{majumder2021move2hear}.  Our work also leverages state-of-the-art AV simulations for learning, but our objective and models are entirely different.  Rather than train virtual robots to move intelligently, our aim is to clean reverberant audio for better speech analysis.

\noindent\textbf{Audio-visual learning from video.}
Multi-modal video understanding has experienced a resurgence of work in the vision, audio, and machine learning literature in recent years.  This includes exciting advances in self-supervised cross-modal feature learning from video~\cite{morgado-spatial-nips2020,lorenzo-nips2020,korbar-nips2018,visual-echoes}, localizing objects in video with both sight and sound~\cite{hu-localize-nips2020}, and audio-visual source separation~\cite{ephrat2018looking,owens2018audio,gao2018objectSounds,zhao2019som,Afouras20audio-visual-objects,visual-voice}.  None of these methods address speech deverberation.

\vspace*{-0.1in}
\section{The Audio-Visual Dereverberation Task}
\vspace*{-0.1in}

We introduce the novel task of \textit{audio-visual dereverberation}. In this task, a speaker (or other sound source) and a listener are situated in a 3D environment, such as the interior of a house. The speaker---whose location is unknown to the listener---produces a speech waveform $A_s$. A superposition of the direct sound and the reverb is captured by the listener, denoted $A_r$. The reverberant speech $A_r$ can be modeled as the convolution of the anechoic source waveform $A_s$ with the room impulse response (RIR) $R$, i.e. $A_r(t) = A_s(t) * R(t)$~\cite{neely_allen_1979}. $R$ is a function of the environment's geometry, the materials that make up the environment, and the relative positioning of the speaker and the listener. It is possible in principle to measure the RIR $R$ for a real-world environment, but doing so can be impractical when the source and listener are able to move around or must cope with different environments. Furthermore, in the common scenario where we want to process video captured in environments to which we have no physical access, measuring the RIR is simply impossible.

Crucially to our task, we consider an alternative source of information about the environment: vision. We assume the listener has an RGB-D observation of its surroundings, obtained from a RGB-D camera or an RGB camera coupled with single-image depth estimation~\cite{Wichern_depth,monodepth2}.
Intuitively,
we should be able to leverage the information about the environment's geometry and material composition that is implicit in the visual stream---\KG{as well as} the location of the speaker (if visible)---to estimate its reverberant characteristics. 
\KG{We anticipate that these cues can}
inform an estimate of the room acoustics, and thus the clean source waveform. Given the RGB $I_r$ and depth image $I_d$ captured by the listener from its current vantage point, the task is to 
predict the source waveform $A_s$ \KG{from the images and reverberant audio}: $\hat{A}_s(t) = f_p([I_r, I_d, A_r(t)])$. This setting represents common real-world scenarios previously discussed, and poses new challenges for speech enhancement and recognition. 

\section{Dataset Curation}\label{sec:dataset} 
For the proposed task, obtaining the right training data is itself a challenge. Existing video data contains reverberant audio but lacks the ground truth anechoic 
audio signal, and existing RIR datasets~\cite{jeub_binaural,jeub_air,open_air} do not have images paired with the microphone position. 
We introduce both real and simulated datasets to enable reproducible research on audio-visual deverberation.

\noindent\textbf{3D environments and acoustic simulator.} 
First we introduce a large-scale dataset in which we couple real-world visual environments with state-of-the-art audio simulations accurately capturing the environments' spatial effects on real samples of recorded speech.  We want our dataset to allow control of a variety of physical environments, the positions of the listener/camera and sources, and the speech content of the sources---all while maintaining both the observed reverberant $A_r(t)$ and ground truth anechoic $A_s(t)$ sounds.  To this end, we leverage the audio-visual simulator SoundSpaces~\cite{chen_soundspaces_2020}, which
provides precomputed RIRs $R(t)$ on a uniform grid of resolution 1 m for the real-world environment scans in Replica~\cite{straub2019replica} and Matterport3D~\cite{Matterport3D}.
We use 82 Matterport environments due to their greater scale and complexity; each environment \KG{has multiple rooms} spanning on average 517 m$^2$. 

\begin{figure}[t]
    \centering
    \includegraphics[width=\linewidth]{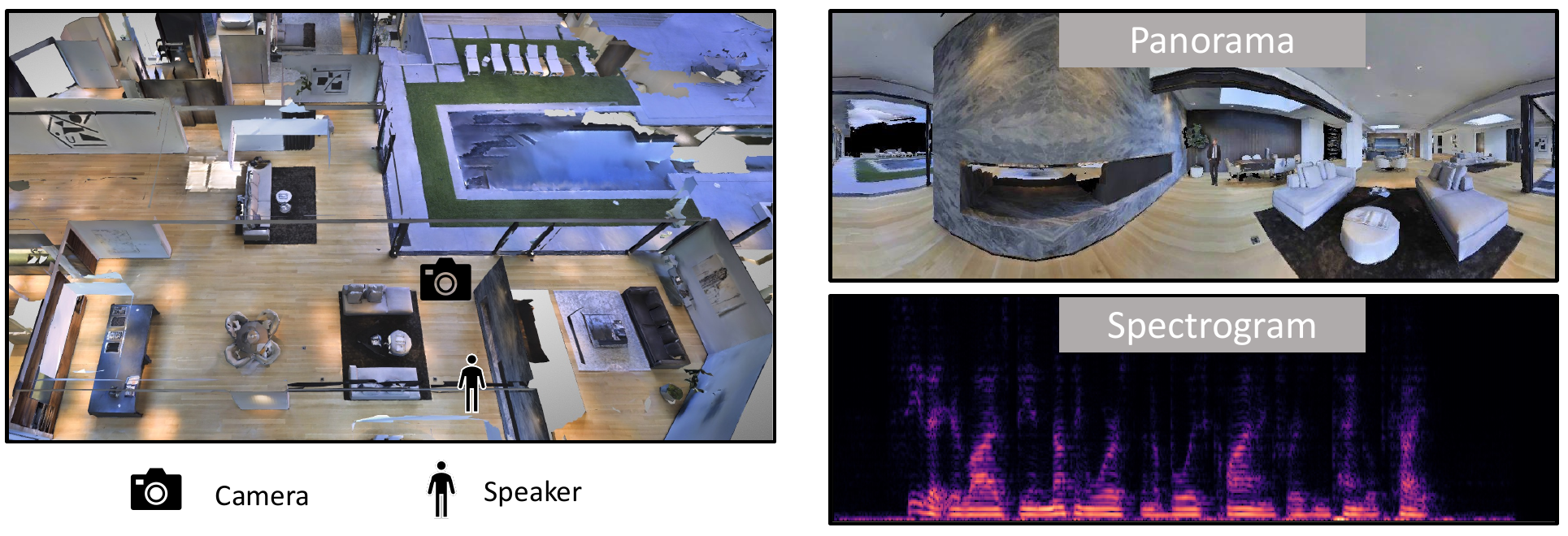}
    \vspace*{-0.15in}
    \caption{\small{\CA{Audio-visual rendering \KG{for a Matterport environment}. Left: bird's-eye view of the 3D environment.  Right: panorama image rendered at the camera location and the corresponding received spectrogram.}}}
    \label{fig:dataset}
    \vspace{-0.1in}
\end{figure}

\noindent\textbf{SoundSpaces-Speech.} We extend SoundSpaces to construct reverberant speech. 
As the source speech corpus we use LibriSpeech~\cite{librispeech}, which contains 1,000 hours of 16kHz read English speech from audio books, and is widely used in the speech recognition literature. 
We train our models with the train-clean-360 split, and use the dev-clean and test-clean sets for validation and test splits, respectively.
Note that these splits have non-overlapping speaker identities. Similarly, we use the standard disjoint train/val/test splits for the Matterport 3D visual environments~\cite{chen_soundspaces_2020}.  Thus, neither the houses nor speaker voices observed at test time are ever observed during training.

For each source utterance, we randomly sample a source-receiver location pair in a random environment, then convolve the speech waveform $A_s(t)$ with the associated SoundSpaces RIR $R(t)$ to obtain the reverberant $A_r(t)$. To augment the visual scene, we insert a 3D humanoid 
of the same gender as the real speaker at the speaker location and  render RGB and depth images at the listener location. We consider two types of image: panorama and normal field of view (FoV).
For the panorama image, we stitch 18 images each having a horizontal FoV of 20 degrees, for a full image resolution of $192\times756$.
For the normal FoV, we render images with a 80 degree FoV, at a resolution of $384\times256$.
While the panorama gives a fuller view of the environment and thus should allow the model to better estimate the room acoustics, the normal FoV is more 
common in existing video and thus will facilitate  our model's transfer to real data. \CA{See Fig.~\ref{fig:dataset}.} 
We generate 49,430/2,700/2,600 such samples for the train/val/test splits, respectively. See Supp.~materials for \KG{examples and} details.

\noindent\textbf{Real data collection.}
To explore whether models trained in simulation can also work in the real world, we also collect a set of real images and speech recordings while preserving the ground truth anechoic audio. 

To collect image data, we use \CA{an iPhone 11 camera to capture a panoramic RGB image 
and a monocular depth estimation algorithm~\cite{monodepth2} to generate the corresponding depth image.}

To record the audio, we use a ZYLIA ZM-1 microphone. We place \CA{both the camera and microphone} at the same height (1.5m) as the SoundSpaces RIRs.
For the source speech, we play utterances from the LibriSpeech test set through a loudspeaker held by a person facing the camera. We collect data from varying environments, including auditoriums, meeting rooms, atriums, corridors, and classrooms. For each environment, we vary the speaker location from near-field to mid-field to far-field. For each location, we play around 10 utterances. 
During data collection, the microphone also records ambient sounds like people chatting, door opening, AC humming, etc. In total, we collect 200 recordings. 
\ISnew{Code and data are available at \url{https://github.com/facebookresearch/learning-audio-visual-dereverberation}.}
\begin{figure*}[ht!]
    \centering
    \includegraphics[width=\textwidth]{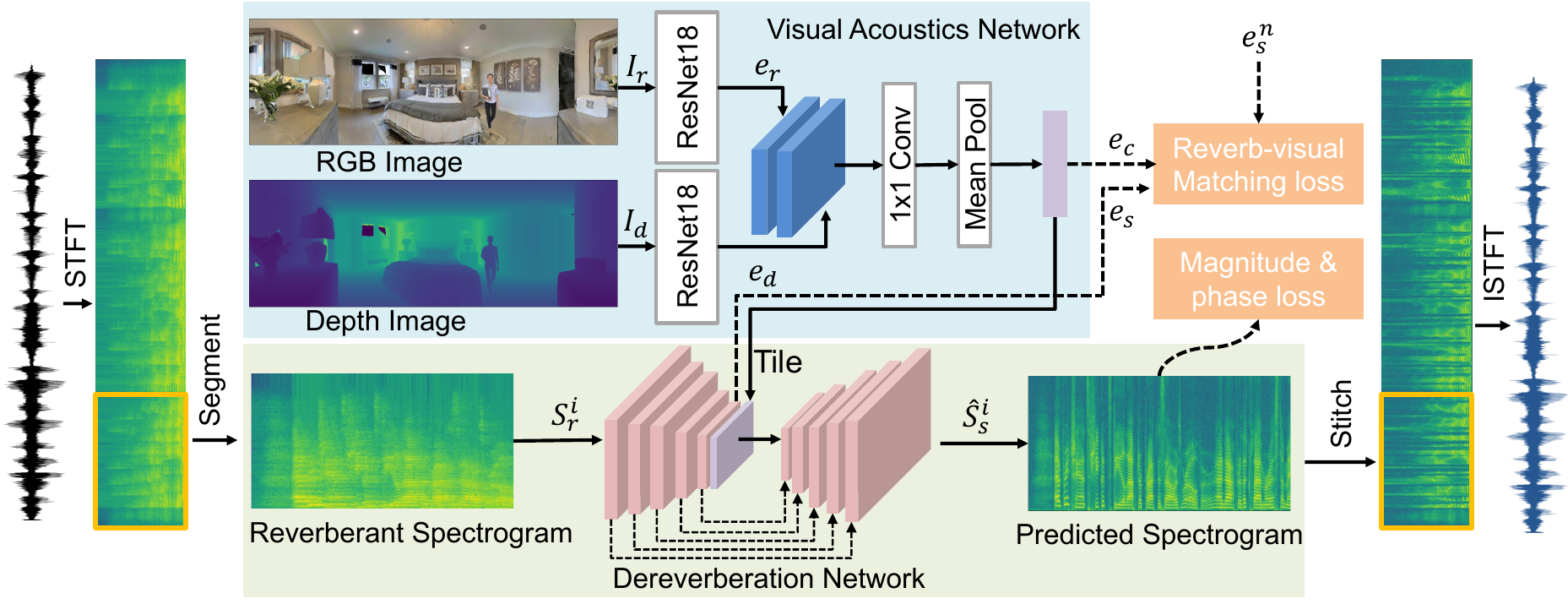}
    \caption{VIDA model architecture. We convert the input speech to a spectrogram and use overlapping sliding windows to obtain 2.56 second segments. For visual inputs, we use separate ResNet18 networks to extract features $e_r$ and $e_d$, which are fused to obtain $e_c$. We feed the spectrogram segment $S_r^i$ to a UNet encoder, tile and concatenate $e_c$ with the encoder's output, then use the UNet decoder to predict the clean spectrogram $\hat{S}_s^i$. During inference, we stitch the predicted spectrogams back into a full spectrogram and use Griffin-Lim~\cite{griffin} to 
    \KG{reconstruct the output dereverberated} waveform.}
    \label{fig:model}
    \vspace*{-0.1in}
\end{figure*}

\section{Approach}
We propose the Visually-Informed Dereverberation of Audio (VIDA) model, which leverages visual cues to learn representations of the environmental acoustics and 
\KG{sound source} locations to dereverberate \KG{audio}. 
\KG{While our model is agnostic to the audio source type, we focus on speech due to the importance of dereverberating speech for downstream analysis.}
VIDA consists of two main components (Figure~\ref{fig:model}): 1) a Visual Acoustics Network (VAN), which learns to map RGB-D images of the environment to features useful for dereverberation, and 2) the dereverberation module itself, which is based on a UNet encoder-decoder architecture. The UNet encoder takes as input a reverberant spectrogram, while the decoder takes the encoder's output along with the visual dereverberation features produced by the VAN and reconstructs a dereverberated version of the \KG{audio}. 

\noindent\textbf{Visual Acoustics Network.} 
Visual observations of a scene reveal information about room acoustics, including room geometry, materials, object locations, and the speaker position. We \KG{devise} 
the VAN to capture all these cues into a latent embedding vector, which is subsequently used to remove reverb.
This network takes as its input an RGB image $I_r$ and a depth image $I_d$, captured from the listener's current position within the environment. The depth image contains information about the geometry of the environment and arrangement of objects, while the RGB image contains more information about their material composition. To better model these different information sources, we use two separate ResNet18~\cite{resnet18} networks to extract their features, i.e. $e_r = f_r(I_r)$ and $e_d = f_d(I_d)$. We concatenate $e_r$ and $e_d$ channel-wise and feed the result to a 1x1 convolution layer $f_c(\cdot)$ to reduce the number of total channels to 512 \KG{followed by} a subsequent pooling layer $f_l(\cdot)$ to reduce the spatial dimension, resulting in the output vector $e_c = f_l(f_c([e_r;e_d]))$.  

\noindent\textbf{Dereverberation Network.}
To recover the clean speech audio, we use the UNet~\cite{unet} architecture, a fully convolutional network often used for image segmentation. We first use the Short-Time Fourier Transform (STFT) to convert the reverberant input audio $A_r$ to a complex spectrogram $S_r$. We treat $S_r$ as a 2-dimensional, 2-channel image, where the horizontal dimension represents time, the vertical dimension represents frequency, and the two channels represent the log-magnitude and phase angle. Our UNet takes spectrograms of a fixed size of $256 \times 256$ as input, but in general the duration of the speech audio we wish to dereverberate will be variable. Therefore, the model processes the full input spectrogram using a series of overlapping, sliding windows. Specifically, we segment the spectrogram along the time dimension into a sequence of fixed-size chunks $S_r^{seg} = \{S_r^1, S_r^2, ..., S_r^n\}$ using a sliding window of length $s$ frames and 50\% overlap between consecutive windows to avoid boundary artifacts. To derive the ground-truth target spectrograms used in training, we perform the exact same segmentation operation on the clean source audio $A_s$ to obtain $S_s^{seg} = \{S_s^1, S_s^2, ..., S_s^n\}$. 

During training, when a particular waveform $S_r$ is selected for inclusion in a data batch, we randomly sample one of its segments $S_r^i$ to be the input to the model, and choose the corresponding $S_s^i$ as the target. We first compute the output of the VAN, $e_c$, for the environment image associated with $S_r$. Next, $S_r^i$ is fed to the UNet's encoder to extract the intermediate feature map $e_s = f_{enc}(S_r^i)$. We then spatially tile and concatenate $e_c$ channel-wise with $e_s$, and feed the fused features to the UNet decoder, which predicts the source spectrogram segment $\hat{S}_s^i = f_{dec}([e_s, e_c])$. 

\noindent\textbf{Spectrogram prediction loss.}
The primary loss function we use to train our model is the Mean-Squared Error (MSE) between the predicted and ground-truth spectrograms, treating the magnitude and phase separately. For a given predicted spectrogram segment $\hat{S}_s^i$, let $\hat{M}_s^i$ denote the predicted log-magnitude spectrogram, $\hat{P}_s^i$ denote the predicted phase spectrogram, and $M_s^i$ and $P_s^i$ denote the respective ground-truth magnitude and phase spectrograms. We define the magnitude loss as:
\begin{equation*}
L_{magnitude} = ||M_s^i - \hat{M}_s^i ||_2.
\end{equation*}
 To address the issue of phase wraparound, we map the phase angle to its corresponding rectangular coordinates on the unit circle and then compute the loss for the phase:
\begin{equation*}
    L_{phase} = ||\sin(P_s^i) - \sin(\hat{P}_s^i)||_2 + ||\cos(P_s^i) - \cos(\hat{P}_s^i)||_2.
\end{equation*}

\noindent\textbf{Reverb-visual matching loss.} To reinforce the consistency between the visually-inferred room acoustics and the reverberation characteristics learned by the UNet encoder, we also employ a contrastive reverb-visual matching loss:
\begin{multline*}
    L_{matching}(e_c, e_s, e_s^n) = \max\{d(f_n(e_c), f_n(e_s)) \\
    - d(f_n(e_c), f_n(e_s^n)) + m, 0\}.
\end{multline*}
Here, $d(x, y)$ represents L2 distance, $f_n(\cdot)$ applies L2 normalization, 
$m$ is a margin, and $e_s^n$ is a different speech embedding sampled from the same data batch. This loss forces the embeddings output by the VAN and the UNet encoder to be consistent, which we empirically show to be beneficial.

\noindent\textbf{Training.}
Our overall training objective (for a single training example) is as follows:
\begin{equation*}
    L_{total} = L_{magnitude} + \lambda_1 L_{phase} + \lambda_2 L_{matching},
\end{equation*}
where $\lambda_1$ and $\lambda_2$ are weighting factors for the phase and matching losses. 
To augment the data, we further choose to rotate the image view for a random angle for each input during training. This is possible because our audio recording is omni-directional and is independent of camera pose. This data augmentation strategy prevents the model from overfitting; without it our model fails to converge. It creates a one-to-many mapping between reverb and views, forcing the model to learn a viewpoint-invariant representation of the room acoustics.

\noindent\textbf{Testing.} At test time, we wish to re-synthesize the entire clean waveform instead of a single fixed-length segment. In this case, we feed all of the segments for a waveform $S_r$ into the model and temporally concatenate all of the output segments. Because consecutive segments overlap by 50\%, during the concatenation step we only retain the middle 50\% of the frames from each segment and discard the rest. Finally, to re-synthesize the waveform we use the Griffin-Lim algorithm~\cite{griffin} to iteratively improve the predicted phase for 30 iterations, which we find works better than directly using the predicted phase or using Griffin-Lim with a randomly initialized phase.

\section{Experiments}
We evaluate our model by dereverberating speech for three downstream tasks: speech enhancement \KG{(SE)}, automatic speech recognition \KG{(ASR)}, and speaker verification \KG{(SV)}. We evaluate using both 
real scanned Matterport3D environments with simulated audio as well as real-world data collected with a camera and mic.  Please see Supp.~for all hyperparameter settings and data preprocessing details.

\noindent\textbf{Evaluation tasks and metrics.} 
We report the standard metrics Perceptual Evaluation of Speech Quality (PESQ) ~\cite{rix_perceptual_2001}, Word Error Rate (WER), and Equal Error Rate (EER) for the three tasks, respectively. For ASR and SV, we use pretrained models from the SpeechBrain~\cite{SB2021} toolkit.  We evaluate these models off-the-shelf on our (de)reverberated version of the LibriSpeech test-clean set, and also explore finetuning the model on the (de)reverberated LibriSpeech train-clean-360 data to ensure all models have exposure to reverberant speech when training. For  speaker verification, \CAn{we construct a set of 80k sampled utterance pairs consisting of different rooms, mic placements, and genders to account for session variability, similar to \cite{richey2018voices}.} Please see Supp. for more details.

\noindent\textbf{Baseline models.} 
In addition to evaluating the the clean and reverberant audio (with no enhancement), we compare against multiple baseline dereverberation models:
\begin{enumerate}[leftmargin=*] 
    \item MetricGAN+~\cite{fu_metricgan_2021}: a recently proposed state-of-the-art model for speech enhancement; we use the public implementation from SpeechBrain~\cite{SB2021}, trained on our dataset. Following the original paper, we optimize for PESQ during training, then choose the best-performing model snapshot (on the validation data) specific to each of our downstream tasks. 
    \item \CAn{HiFi-GAN~\cite{su_hifi-gan_2020}: a recent model for denoising and dereverberation. We use this implementation: \url{https://github.com/rishikksh20/hifigan-denoiser}.}
    \item WPE~\cite{wpe}: A statistical speech dereverberation model that is commonly used for comparison.
\end{enumerate}
We emphasize that all baselines are \textit{audio-only} models, as opposed to our proposed \textit{audio-visual} model. 
Our multimodal dereverberation technique could extend to work in conjunction with other newly-proposed audio-only models, \KG{i.e., ongoing architecture advances are orthogonal to our idea.}

\begin{table}[t!]\hspace*{-0.2in}
\setlength{\tabcolsep}{4pt}
\centering
\begin{tabular}{ c| c | c c | c c}

 \toprule
                                                & \multicolumn{1}{c|}{\textit{SE}}
                                                & \multicolumn{2}{c|}{\textit{ASR}} & \multicolumn{2}{c}{\textit{SV}} \\
                                                & {PESQ } & {WER } & {FT  } & {EER } & {FT}\\
 \midrule
    \CAr{Anechoic}   \KG{(Upper bound)}          & 4.64 & 2.50  &  2.50 & 1.62 & 1.62\\
    \midrule
    Reverberant                         & 1.54 & 8.86  &  4.62 & 4.69 & 4.57 \\
    MetricGAN+~\cite{fu_metricgan_2021} & 2.33 & 7.49 &  4.86 & 4.67 & 2.75 \\
    HiFi-GAN~\cite{su_hifi-gan_2020} & 1.83 & 9.31 &  5.59 & 4.30 & 2.49 \\
    WPE~\cite{wpe}                      & 1.63 & 8.18  &  4.30 &  5.19 & 4.48 \\
\midrule
    VIDA w/o VAN                        & 2.32 & 4.92  & 3.76 & 4.67 & 2.61\\
    VIDA w/ normal FoV                  & 2.33 & 4.85  & 3.73 & 4.53 & 2.79 \\
    VIDA w/o matching loss              & \bfseries 2.38 & 4.59  & 3.72 & 4.02 & 2.62 \\
    VIDA w/o human mesh                 & 2.31 & 4.57 & 3.72 & 4.00 & 2.52 \\
    VIDA w/ random image                & 2.34 & 4.94  & 3.82 & 4.70 & 2.48 \\
    VIDA                                & 2.37 & \bfseries 4.44  & \bfseries 3.66  & \bfseries 3.97 & \bfseries 2.40 \\
 \bottomrule
\end{tabular}
\vspace*{-0.1in}
\caption{Results on \CAcut{multiple speech analysis tasks, evaluated on the} LibriSpeech test-clean set that is 
reverberated with our environmental simulator (with the exception of the ``\CAr{Anechoic} (Upper bound)'' setting, which is evaluated on the original audio). \KGs{FT} 
refers to tests where the 
models are finetuned with the audio-enhanced data. 
\CAcut{\CA{The relative improvement compared to Reverberant is included in parentheses.}}
}
\vspace*{-0.1in}
\label{tab:main_results}
\end{table}

\noindent\textbf{Results \CAis{on SoundSpaces-Speech}.} 
Table~\ref{tab:main_results} shows the results for all models on SE, ASR, and SV. 
\KG{First, since existing methods report results on \CAr{anechoic} audio, we note the pretrained SpeechBrain model applied to \CAr{anechoic} audio (first row) yields errors competitive with the SoTA~\cite{conformer}, meaning we have a solid experimental testbed.}
Comparing the results on \CAr{anechoic} vs.~reverberated speech, we see that reverberation significantly degrades performance on all tasks. Our VIDA model outperforms all other models, and by a large margin on the ASR and SV tasks.\CAcut{without finetuning, we achieve absolute improvements of 0.04 PESQ (1.71\% relative improvement), 0.48\% WER (9.75\% relative improvement), and 0.68\% EER (14.56\% relative improvement) over the \emph{best baseline} in each case (which happens to be the audio-only version of VIDA for both the ASR and SV tasks).} \KGs{The results are statistically significant according to a paired t-test.} 
After finetuning the ASR model, the gain is still largely preserved at 0.64\% WER (14.88\% relative), although it is important to note that finetuning downstream models on enhanced speech is not always feasible, \KGiclr{e.g., if using off-the-shelf ASR}. 
Our results demonstrate that learning the acoustic properties\CAcut{of an environment} from visual signals is very helpful for dereverberating speech, enabling the model to leverage information 
unavailable in the audio alone.

\noindent\textbf{Ablations.}
\ISnew{To study how much VIDA leverages visual signals, we ablate the visual network VAN (audio-only). Table~\ref{tab:main_results} shows the results. All performance degrades significantly, showing that visual acoustic features are helpful for dereveberation.}
To understand how well VIDA works with a normal field-of-view (FoV) camera, we replace the panorama image input with a FoV of 80 degrees randomly sampled from the current view. All metrics drop compared to using a panorama, \CAis{as expected}.\CAcut{This is expected, because \KG{the model is limited in what it can see} with a narrower field of view; 
\KG{the inferred room acoustics are impaired by} not seeing the full environment or missing where the speaker is.}  Compared to the audio-only \ISnew{ablation}, however, VIDA still performs better;  even a partial view of the environment helps the model understand the scene and dereverberate the audio.
\KG{Next, we} ablate the proposed reverb-visual matching loss (``w/o matching loss").  Without it, VIDA's performance \KG{declines} on all metrics. This shows by forcing the visual feature to agree with the reverberation feature, our model learns a better representation of room acoustics. 
To examine how much the model leverages the human speaker cues \CAn{and uses the visual scene}, 
we evaluate 
\KGcr{VIDA} on the same test data but with the 3D humanoid removed \KG{(``w/o human mesh")}
\CAr{or train VIDA with random images (``w/ random image") and \KGcr{re-}evaluate}.
All three metrics become worse.
This shows our model pays attention to \CAn{both} the \CAn{presence} of the human speaker \CAn{and the scene geometry} to 
\KGcr{better} \KG{anticipate reverberation}.

\begin{table}[t!]\small
\setlength{\tabcolsep}{5pt}
\centering
\begin{tabular}{ c| c | c | c}
 \toprule
                                                & \multicolumn{1}{c|}{\textit{SE}} & \multicolumn{1}{c|}{\textit{ASR}} & \multicolumn{1}{c}{\textit{SV}}\\
                                                & {PESQ} & {WER} & {EER}\\
 \midrule
    \CAr{Anechoic}   \KG{(Upper bound)}                 &   4.64   & 2.52  & 1.42\\
    \midrule
    Reverberant                                & 1.22 &  18.39  & 3.91\\
    MetricGAN+~\cite{fu_metricgan_2021}        & \bfseries 1.62 & 21.42 & 5.70\\
    HiFi-GAN~\cite{su_hifi-gan_2020}           & 1.33 & 24.05 & 5.21\\
    \midrule
    VIDA w/o VAN                               & 1.41 &   15.18 & 4.24 \\
    VIDA w/ normal FoV                   & 1.44  &  14.71  & 3.79  \\
    VIDA                                & 1.49   &  \bfseries 13.02  & \bfseries 3.75 \\
 \bottomrule
\end{tabular}
\vspace*{-0.1in}
\caption{Results on real data \KG{demonstrating sim2real transfer.  
} 
}
\label{tab:real_data}
\vspace*{-0.1in}
\end{table}

\begin{table}[t!]\small
\setlength{\tabcolsep}{3pt}
\centering
\begin{tabular}{ c| c| c| c| c }
 \toprule
 &  {Atrium} & {Conf. Room} & {Classroom} & {Corridor} \\
 \midrule
    Near-field &  14.1 / \textbf{9.0}    & \textbf{5.0} / 6.5 & 6.1 / \textbf{5.3} & 2.2 / \textbf{1.8}\\
    Mid-field  & 21.8 / \textbf{18.9}    &  \textbf{7.7} / \textbf{7.7} & 2.6 / \textbf{1.5} &  7.3 / \textbf{4.4}\\
    Far-field  & 52.4 / \textbf{50.5}    & 22.0 / \textbf{6.7} & \textbf{5.9} / 6.8 & 25.2 / \textbf{21.1} \\
 \bottomrule
\end{tabular}
\vspace*{-0.1in}
\caption{Breakdown of word error rate (WER) for \ISnew{VIDA without and with VAN} \KG{on real test data}.   
}
\label{tab:real_data_breakdown}
\vspace*{-0.1in}
\end{table}

\noindent\textbf{Results on real data.} Next, we deploy our model in the real world. We use \KG{all} models trained in simulation to dereverberate the real-world dataset (cf.~Sec.~\ref{sec:dataset}) before using the finetuned ASR/SV models to evaluate the enhanced speech.
Table~\ref{tab:real_data} shows the results of all models on real data. Reverberation does more damage to the WER compared to in simulation. Although MetricGAN+~\cite{fu_metricgan_2021} has 
better PESQ, it has a \KG{weak} WER score. Our VIDA model \KG{again} outperforms all baselines on ASR and SV. 
This demonstrates the realism of the simulation and the capability of our model to transfer to real-world data, \KG{a promising step for VIDA's wider applicability.}

Table~\ref{tab:real_data_breakdown} breaks down the ASR performance for \ISnew{VIDA without and with VAN} by environment type and speaker distance.
The atrium is quite reverberant due to the large space.\CAcut{Although the auditorium is similarly large, the space is designed to reduce reverberation and thus both models have lower WER.} The conference room and the classroom have smaller sizes and are comparatively less reverberant. The corridor only becomes reverberant when the speaker is far away. 
\CA{VIDA outperforms VIDA w/o VAN in most cases, especially in highly reverberant ones.}

\begin{figure}[t]
\centering
\subcaptionbox{Audio t-SNE\label{fig:tsne_audio_distance}}
{\includegraphics[width=0.235\textwidth]{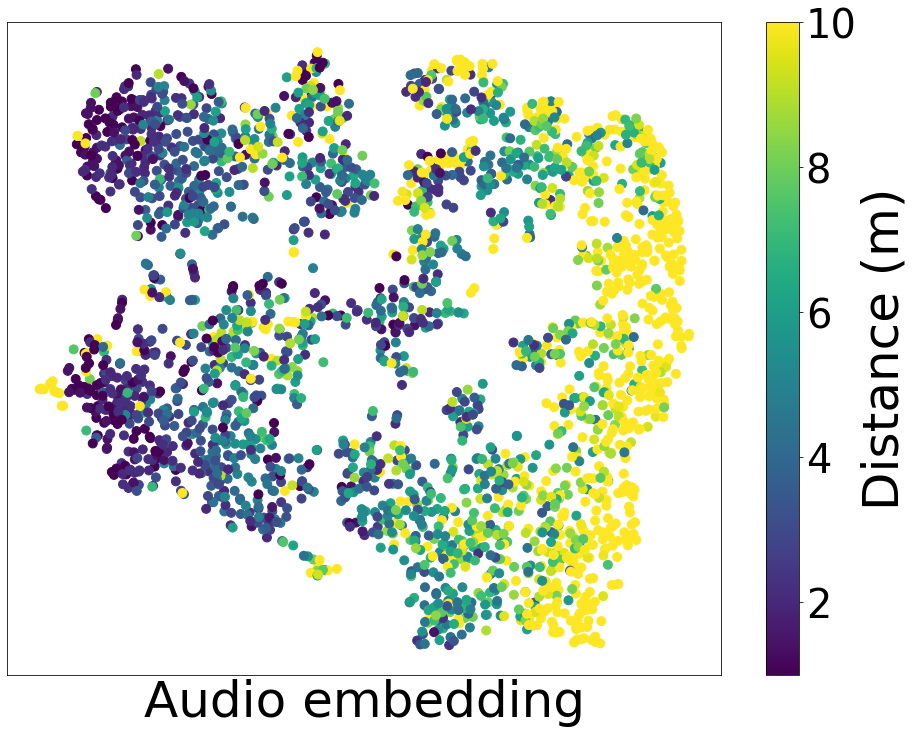}}
\subcaptionbox{Visual t-SNE\label{fig:tsne_visual_rt60}}
{\includegraphics[width=0.235\textwidth]{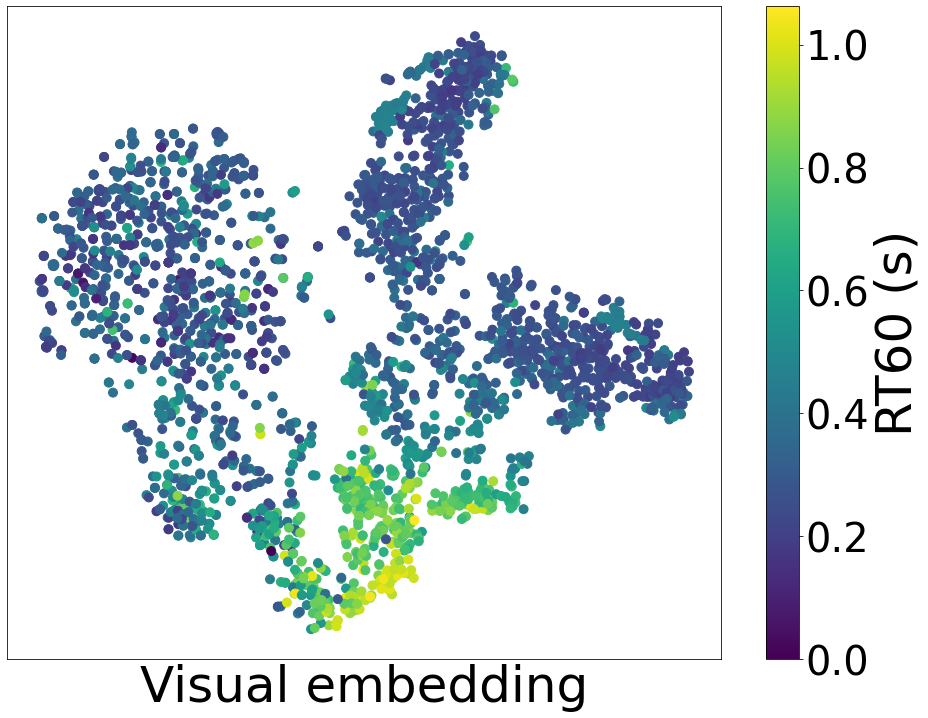}}
\caption{
t-SNE of audio and visual features colored by the distance to the speaker \KG{(c)} and RT60 \KG{(d)}. 
}
\vspace*{-0.05in}
\label{fig:distribution}
\end{figure}

\noindent\textbf{Analyzing learned features.} Figure~\ref{fig:tsne_audio_distance} and~\ref{fig:tsne_visual_rt60} analyze our model's learned audio and visual features via 2D t-SNE projections~\cite{tsne}. For each sample, we color the point according to either (c) the ground truth distance between the camera/microphone and the human speaker or (d) the reverberation time for the \CA{audio signal} to decay by 60 dB (known as the RT60).  Neither of these variables are available to our model during training, yet when learning to perform deverberation, our model exposes these high-level properties relevant to the audio-visual task.  Consistent with the quantitative results above, this analysis shows how our model captures elements of the visual scene, room geometry, and speaker location that are valuable to proper dereverberation. 

\begin{figure}[t]
    \centering
    \includegraphics[width=\linewidth]{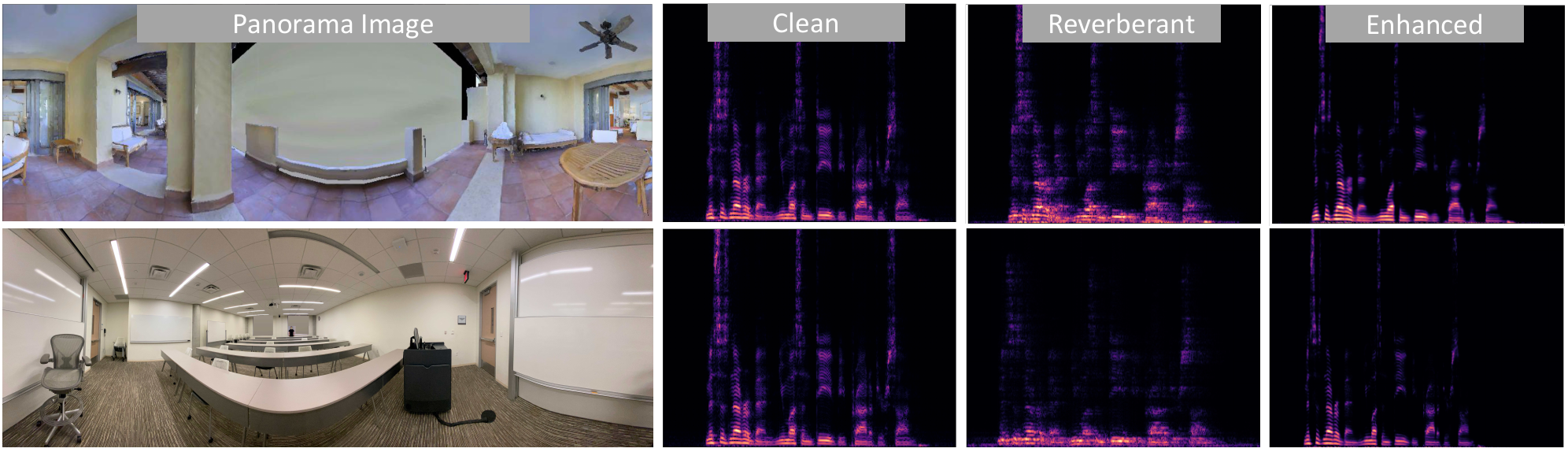}
    \caption{Example input images, clean spectrograms, reverberant spectograms and spectrograms dereverberated by VIDA \KG{(top is from a scan, bottom is a real pano)}. \CA{The speaker is out of view in the first case and distant in the second case (back of the classroom).}
    Though both received audio inputs are quite reverberant, our model successfully removes the reverb and restores the clean source speech.
    }
    \label{fig:spectrogram_comparison}
\vspace*{-0.05in}
\end{figure}

\noindent\textbf{Qualitative examples.} Figure~\ref{fig:spectrogram_comparison} shows \CA{a simulated and real-world example}. As we can see, the reverberant spectrogram is much blurrier compared to the clean spectrogram, while our predicted spectrogram removes those reverberations by leveraging the visual cues of room acoustics.

\section{Conclusion}
We introduced the novel task of audio-visual dereverberation.  The proposed VIDA approach learns to remove reverb by attending to both the audio and visual streams, recovering valuable signals about room geometry, materials, and speaker locations from visual encodings of the environment.  In support of this task, we develop a large-scale dataset providing realistic, spatially registered observations of speech and 3D environments. VIDA successfully dereverberates novel voices in novel environments more accurately than an array of baselines, improving 
multiple downstream tasks.\CAcut{Trained in a simulator, our model also shows promise enhancing speech in real-world data, \KG{though more work is needed to bring down the absolute error rates.} }
In future work, we will \KG{explore temporal models for dereverberation with \CAn{real-world} video\CAcut{and introduce active perception for camera control}.}

\noindent\textbf{Acknowledgements:} UT Austin is supported in part by the IFML NSF AI Institute and NSF CCRI.  KG is paid as a researcher at Meta.
\clearpage

\bibliographystyle{IEEEbib}
\bibliography{ref}

\begin{thebibliography}{10}

\bibitem{reverb-challenge}
Keisuke Kinoshita, Marc Delcroix, Sharon Gannot, Emanuël Habets, Reinhold
  Haeb-Umbach, Walter Kellermann, Volker Leutnant, Roland Maas, Tomohiro
  Nakatani, Bhiksha Raj, Armin Sehr, and Takuya Yoshioka,
\newblock ``A summary of the reverb challenge: State-of-the-art and remaining
  challenges in reverberant speech processing research,''
\newblock {\em EURASIP Journal on Advances in Signal Processing}, 2016.

\bibitem{zhao_monaural_2020}
Yan Zhao, DeLiang Wang, Buye Xu, and Tao Zhang,
\newblock ``Monaural speech dereverberation using temporal convolutional
  networks with self attention,''
\newblock 2020.

\bibitem{szoke_2019}
Igor Sz{\"o}ke, Miroslav Sk{\'{a}}cel, Ladislav Mo{\v{s}}ner, Jakub Paliesek,
  and Jan {\v{C}}ernock{\'{y}},
\newblock ``Building and evaluation of a real room impulse response dataset,''
\newblock {\em IEEE Journal of Selected Topics in Signal Processing}, 2019.

\bibitem{han_learning_2015}
Kun Han, Yuxuan Wang, DeLiang Wang, William~S. Woods, Ivo Merks, and Tao Zhang,
\newblock ``Learning spectral mapping for speech dereverberation and
  denoising,''
\newblock in {\em IEEE Trans Audio Speech Lang Process}, 2015.

\bibitem{wu_end--end_2017}
Bo~Wu, Kehuang Li, Fengpei Ge, Zhen Huang, Minglei Yang, Sabato~Marco
  Siniscalchi, and Chin-Hui Lee,
\newblock ``An end-to-end deep learning approach to simultaneous speech
  dereverberation and acoustic modeling for robust speech recognition,''
\newblock in {\em {IEEE} Journal of Selected Topics in Signal Processing},
  2017.

\bibitem{ernst_speech_2019}
Ori Ernst, Shlomo~E. Chazan, Sharon Gannot, and Jacob Goldberger,
\newblock ``Speech dereverberation using fully convolutional networks,''
\newblock in {\em 2018 26th European Signal Processing Conference (EUSIPCO)},
  2018.

\bibitem{giri_improving_2015}
``Improving speech recognition in reverberation using a room-aware deep neural
  network and multi-task learning,''
\newblock in {\em 2015 {IEEE} International Conference on Acoustics, Speech and
  Signal Processing ({ICASSP})}.

\bibitem{xvector}
David Snyder, Daniel Garcia-Romero, Gregory Sell, Daniel Povey, and Sanjeev
  Khudanpur,
\newblock ``X-vectors: Robust {DNN} embeddings for speaker recognition,''
\newblock in {\em 2018 IEEE International Conference on Acoustics, Speech and
  Signal Processing (ICASSP)}. IEEE, 2018, pp. 5329--5333.

\bibitem{wu2016}
Bo~Wu, Kehuang Li, Minglei Yang, and Chin-Hui Lee,
\newblock ``A reverberation-time-aware approach to speech dereverberation based
  on deep neural networks,''
\newblock in {\em IEEE Trans Audio Speech Lang Process}, 2016.

\bibitem{zhao2019}
Yan Zhao, Zhong-Qiu Wang, and DeLiang Wang,
\newblock ``Two-stage deep learning for noisy-reverberant speech enhancement,''
\newblock {\em IEEE Trans Audio Speech Lang Process}, 2019.

\bibitem{su_hifi-gan_2020}
Jiaqi Su, Zeyu Jin, and Adam Finkelstein,
\newblock ``{HiFi}-{GAN}: High-fidelity denoising and dereverberation based on
  speech deep features in adversarial networks,''
\newblock in {\em INTERSPEECH}, 2020.

\bibitem{wpe}
Tomohiro Nakatani, Takuya Yoshioka, Keisuke Kinoshita, Masato Miyoshi, and
  Biing-Hwang Juang,
\newblock ``Speech dereverberation based on variance-normalized delayed linear
  prediction,''
\newblock in {\em IEEE Transactions on Audio, Speech, and Language Processing},
  2010.

\bibitem{naylor_gaubitch_2010}
Patrick~A. Naylor and Nikolay~D. Gaubitch,
\newblock {\em Speech Dereverberation},
\newblock Springer Publishing Company, Incorporated, 1st edition, 2010.

\bibitem{fu_metricgan_2019}
Szu-Wei Fu, Chien-Feng Liao, Yu~Tsao, and Shou-De Lin,
\newblock ``{MetricGAN}: Generative adversarial networks based black-box metric
  scores optimization for speech enhancement,''
\newblock in {\em ICML}, 2019.

\bibitem{majumder2022fewshot}
Sagnik Majumder, Changan Chen, Ziad Al-Halah, and Kristen Grauman,
\newblock ``Few-shot audio-visual learning of environment acoustics,''
\newblock in {\em NeurIPS}, 2022.

\bibitem{chen22vam}
Changan Chen, Ruohan Gao, Paul Calamia, and Kristen Grauman,
\newblock ``Visual acoustic matching,''
\newblock in {\em CVPR}, 2022.

\bibitem{luo2022learning}
Andrew Luo, Yilun Du, Michael~J Tarr, Joshua~B Tenenbaum, Antonio Torralba, and
  Chuang Gan,
\newblock ``Learning neural acoustic fields,''
\newblock in {\em NeurIPS}, 2022.

\bibitem{kim2020acoustic}
Hansung Kim, Luca Remaggi, Sam Fowler, Philip~JB Jackson, and Adrian Hilton,
\newblock ``Acoustic room modelling using 360 stereo cameras,''
\newblock {\em IEEE Transactions on Multimedia}, 2020.

\bibitem{chen_soundspaces_2020}
Changan Chen, Unnat Jain, Carl Schissler, Sebastia Vicenc~Amengual Gari, Ziad
  Al-Halah, Vamsi~Krishna Ithapu, Philip Robinson, and Kristen Grauman,
\newblock ``{SoundSpaces}: Audio-visual navigation in 3d environments,''
\newblock in {\em ECCV}, 2020.

\bibitem{librispeech}
Vassil Panayotov, Guoguo Chen, Daniel Povey, and Sanjeev Khudanpur,
\newblock ``Librispeech: an asr corpus based on public domain audio books,''
\newblock in {\em 2015 IEEE international conference on acoustics, speech and
  signal processing (ICASSP)}. IEEE, 2015, pp. 5206--5210.

\bibitem{neely_allen_1979}
{Stephen T.} Neely and {Jont B.} Allen,
\newblock ``Invertibility of a room impulse response,''
\newblock in {\em Journal of the Acoustical Society of America}, 1979.

\bibitem{miyoshi_kaneda_1988}
M.~Miyoshi and Y.~Kaneda,
\newblock ``Inverse filtering of room acoustics,''
\newblock {\em IEEE Transactions on Acoustics, Speech, and Signal Processing},
  1988.

\bibitem{benetsy_speech_enhancement}
Jacob Benesty, Shoji Makino, and Jingdong Chen,
\newblock {\em Speech Enhancement},
\newblock Springer Publishing Company, Incorporated, 1st edition, 2010.

\bibitem{vondrick-denoising-speech-nips2019}
Ruilin Xu, Rundi Wu, Yuko Ishiwaka, Carl Vondrick, and Changxi Zheng,
\newblock ``Listening to sounds of silence for speech denoising,''
\newblock in {\em NeurIPS}, 2019.

\bibitem{hershey2016deep}
John~R Hershey, Zhuo Chen, Jonathan Le~Roux, and Shinji Watanabe,
\newblock ``Deep clustering: Discriminative embeddings for segmentation and
  separation,''
\newblock in {\em ICASSP}, 2016.

\bibitem{stoller2018adversarial}
Daniel Stoller, Sebastian Ewert, and Simon Dixon,
\newblock ``Adversarial semi-supervised audio source separation applied to
  singing voice extraction,''
\newblock in {\em ICASSP}, 2018.

\bibitem{ko2017study}
Tom Ko, Vijayaditya Peddinti, Daniel Povey, Michael~L Seltzer, and Sanjeev
  Khudanpur,
\newblock ``A study on data augmentation of reverberant speech for robust
  speech recognition,''
\newblock in {\em 2017 IEEE International Conference on Acoustics, Speech and
  Signal Processing (ICASSP)}. IEEE, 2017, pp. 5220--5224.

\bibitem{su2020acoustic}
Jiaqi Su, Zeyu Jin, and Adam Finkelstein,
\newblock ``Acoustic matching by embedding impulse responses,''
\newblock in {\em ICASSP 2020-2020 IEEE International Conference on Acoustics,
  Speech and Signal Processing (ICASSP)}. IEEE, 2020, pp. 426--430.

\bibitem{tan_audio-visual_2020}
Ke~Tan, Yong Xu, Shi-Xiong Zhang, Meng Yu, and Dong Yu,
\newblock ``Audio-visual speech separation and dereverberation with a two-stage
  multimodal network,''
\newblock in {\em {IEEE} Journal of Selected Topics in Signal Processing},
  2020.

\bibitem{stan2002comparison}
Guy-Bart Stan, Jean-Jacques Embrechts, and Dominique Archambeau,
\newblock ``Comparison of different impulse response measurement techniques,''
\newblock {\em journal of the audio engineering society}, vol. 50, no. 4, pp.
  249--262, april 2002.

\bibitem{Holters_rir}
Martin Holters, Tobias Corbach, and Udo Zölzer,
\newblock ``Impulse response measurement techniques and their applicability in
  the real world,''
\newblock in {\em Proceedings of the 12th International Conference on Digital
  Audio Effects, DAFx 2009}, 2009.

\bibitem{image_method}
Jont~B. Allen and David~A. Berkley,
\newblock ``Image method for efficiently simulating small‐room acoustics,''
\newblock {\em The Journal of the Acoustical Society of America}, vol. 65, no.
  4, pp. 943--950, 1979.

\bibitem{Murphy_waveguide}
D.T. Murphy, Antti Kelloniemi, Jack Mullen, and Simon Shelley,
\newblock ``Acoustic modeling using the digital waveguide mesh,''
\newblock {\em Signal Processing Magazine, IEEE}, vol. 24, pp. 55 -- 66, 04
  2007.

\bibitem{kon_estimation_2019}
Homare Kon and Hideki Koike,
\newblock ``Estimation of late reverberation characteristics from a single
  two-dimensional environmental image using convolutional neural networks,''
\newblock in {\em Journal of the Audio Engineering Society}, 2019.

\bibitem{singh_image2reverb_2021}
Nikhil Singh, Jeff Mentch, Jerry Ng, Matthew Beveridge, and Iddo Drori,
\newblock ``Image2reverb: Cross-modal reverb impulse response synthesis,''
\newblock {\em arXiv preprint arXiv:2103.14201}, 2021.

\bibitem{morgado-2018}
Pedro Morgado, Nono Vasconcelos, Timothy Langlois, and Oliver Wang,
\newblock ``Self-supervised generation of spatial audio for 360 video,''
\newblock in {\em NeurIPS}, 2018.

\bibitem{25d-visual-sound}
Ruohan Gao and Kristen Grauman,
\newblock ``2.5d visual sound,''
\newblock in {\em CVPR}, 2019.

\bibitem{scene-aware-360}
Dingzeyu Li, Timothy~R Langlois, and Changxi Zheng,
\newblock ``Scene-aware audio for 360 videos,''
\newblock {\em ACM Transactions on Graphics (TOG)}, vol. 37, no. 4, pp. 1--12,
  2018.

\bibitem{jeub_binaural}
Peter~Vary Marco~Jeub, Magnus~Schafer,
\newblock ``A binaural room impulse response database for the evaluation of
  dereverberation algorithms,''
\newblock in {\em Proceedings of International Conference on Digital Signal
  Processing}, 2009.

\bibitem{jeub_air}
Marco Jeub, Magnus Schäfer, Hauke Krüger, Christoph~Matthias Nelke,
  Christophe Beaugeant, and Peter Vary,
\newblock ``Do we need dereverberation for hand-held telephony?,''
\newblock in {\em International Congress on Acoustics}, 2010.

\bibitem{open_air}
Damian~T Murphy and Simon Shelley,
\newblock ``Openair: An interactive auralization web resource and database,''
\newblock in {\em Audio Engineering Society Convention 129}, 2010.

\bibitem{chen22soundspaces2}
Changan Chen, Carl Schissler, Sanchit Garg, Philip Kobernik, Alexander Clegg,
  Paul Calamia, Dhruv Batra, Philip~W Robinson, and Kristen Grauman,
\newblock ``Soundspaces 2.0: A simulation platform for visual-acoustic
  learning,''
\newblock in {\em NeurIPS}, 2022.

\bibitem{gan2019look}
Chuang Gan, Yiwei Zhang, Jiajun Wu, Boqing Gong, and Joshua~B Tenenbaum,
\newblock ``Look, listen, and act: Towards audio-visual embodied navigation,''
\newblock in {\em ICRA}, 2020.

\bibitem{chen_savi_2021}
Changan Chen, Ziad Al-Halah, and Kristen Grauman,
\newblock ``Semantic audio-visual navigation,''
\newblock in {\em CVPR}, 2021.

\bibitem{chen_waypoints_2020}
Changan Chen, Sagnik Majumder, Ziad Al-Halah, Ruohan Gao, Santhosh~Kumar
  Ramakrishnan, and Kristen Grauman,
\newblock ``Learning to set waypoints for audio-visual navigation,''
\newblock in {\em ICLR}, 2021.

\bibitem{dean-curious-nips2020}
Victoria Dean, Shubham Tulsiani, and Abhinav Gupta,
\newblock ``See, hear, explore: Curiosity via audio-visual association,''
\newblock in {\em NeurIPS}, 2020.

\bibitem{majumder2021move2hear}
Sagnik Majumder, Ziad Al-Halah, and Kristen Grauman,
\newblock ``Move2hear: Active audio-visual source separation,''
\newblock in {\em ICCV}, 2021.

\bibitem{morgado-spatial-nips2020}
Pedro Morgado, Yi~Li, and Nuno Vasconcelos,
\newblock ``Learning representations from audio-visual spatial alignment,''
\newblock in {\em NeurIPS}, 2020.

\bibitem{lorenzo-nips2020}
Humam Alwassel, Dhruv Mahajan, Bruno Korbar, Lorenzo Torresani, Bernard Ghanem,
  and Du~Tran,
\newblock ``Self-supervised learning by cross-modal audio-video clustering,''
\newblock in {\em NeurIPS}, 2020.

\bibitem{korbar-nips2018}
Bruno Korbar, Du~Tran, and Lorenzo Torresani,
\newblock ``Cooperative learning of audio and video models from self-supervised
  synchronization,''
\newblock in {\em NeurIPS}, 2018.

\bibitem{visual-echoes}
Ruohan Gao, Changan Chen, Ziad Al-Halah, Carl Schissler, and Kristen Grauman,
\newblock ``Visualechoes: Spatial image representation learning through
  echolocation,''
\newblock in {\em ECCV}, 2020.

\bibitem{hu-localize-nips2020}
Di~Hu, Rui Qian, Minyue Jiang, Xiao Tan, Shilei Wen, Errui Ding, Weiyao Lin,
  and Dejing Dou,
\newblock ``Discriminative sounding objects localization via self-supervised
  audiovisual matching,''
\newblock in {\em NeurIPS}, 2020.

\bibitem{ephrat2018looking}
Ariel Ephrat, Inbar Mosseri, Oran Lang, Tali Dekel, Kevin Wilson, Avinatan
  Hassidim, William~T Freeman, and Michael Rubinstein,
\newblock ``Looking to listen at the cocktail party: A speaker-independent
  audio-visual model for speech separation,''
\newblock in {\em SIGGRAPH}, 2018.

\bibitem{owens2018audio}
Andrew Owens and Alexei~A Efros,
\newblock ``Audio-visual scene analysis with self-supervised multisensory
  features,''
\newblock in {\em ECCV}, 2018.

\bibitem{gao2018objectSounds}
Ruohan Gao, Rogerio Feris, and Kristen Grauman,
\newblock ``Learning to separate object sounds by watching unlabeled video,''
\newblock in {\em ECCV}, 2018.

\bibitem{zhao2019som}
Hang Zhao, Chuang Gan, Wei-Chiu Ma, and Antonio Torralba,
\newblock ``The sound of motions,''
\newblock in {\em ICCV}, 2019.

\bibitem{Afouras20audio-visual-objects}
Triantafyllos Afouras, Andrew Owens, Joon-Son Chung, and Andrew Zisserman,
\newblock ``Self-supervised learning of audio-visual objects from video,''
\newblock in {\em ECCV}, 2020.

\bibitem{visual-voice}
Ruohan Gao and Kristen Grauman,
\newblock ``Visualvoice: Audio-visual speech separation with cross-modal
  consistency,''
\newblock in {\em CVPR}, 2021.

\bibitem{Wichern_depth}
David Eigen, Christian Puhrsch, and Rob Fergus,
\newblock ``Depth map prediction from a single image using a multi-scale deep
  network,''
\newblock in {\em NIPS}, 2014.

\bibitem{monodepth2}
Cl{\'{e}}ment Godard, Oisin {Mac Aodha}, Michael Firman, and Gabriel~J.
  Brostow,
\newblock ``Digging into self-supervised monocular depth prediction,''
\newblock in {\em ICCV}, 2019.

\bibitem{straub2019replica}
Julian Straub, Thomas Whelan, Lingni Ma, Yufan Chen, Erik Wijmans, Simon Green,
  Jakob~J Engel, Raul Mur-Artal, Carl Ren, Shobhit Verma, et~al.,
\newblock ``The replica dataset: A digital replica of indoor spaces,''
\newblock {\em arXiv:1906.05797}, 2019.

\bibitem{Matterport3D}
Angel Chang, Angela Dai, Thomas Funkhouser, Maciej Halber, Matthias Niessner,
  Manolis Savva, Shuran Song, Andy Zeng, and Yinda Zhang,
\newblock ``Matterport3d: Learning from rgb-d data in indoor environments,''
\newblock {\em 3DV}, 2017,
\newblock MatterPort3D dataset license available at:
  \url{http://kaldir.vc.in.tum.de/matterport/MP_TOS.pdf}.

\bibitem{griffin}
Daniel Griffin and Jae Lim,
\newblock ``Signal estimation from modified short-time fourier transform,''
\newblock {\em IEEE Transactions on acoustics, speech, and signal processing},
  vol. 32, no. 2, pp. 236--243, 1984.

\bibitem{resnet18}
Kaiming He, Xiangyu Zhang, Shaoqing Ren, and Jian Sun,
\newblock ``Deep residual learning for image recognition,''
\newblock in {\em CVPR}, 2016, pp. 770--778.

\bibitem{unet}
Olaf Ronneberger, Philipp Fischer, and Thomas Brox,
\newblock ``U-net: Convolutional networks for biomedical image segmentation,''
\newblock in {\em International Conference on Medical image computing and
  computer-assisted intervention}. Springer, 2015, pp. 234--241.

\bibitem{rix_perceptual_2001}
A.~W. Rix, J.~G. Beerends, M.~P. Hollier, and A.~P. Hekstra,
\newblock ``Perceptual evaluation of speech quality ({PESQ})-a new method for
  speech quality assessment of telephone networks and codecs,''
\newblock in {\em 2001 {IEEE} International Conference on Acoustics, Speech,
  and Signal Processing. Proceedings}.

\bibitem{SB2021}
Mirco Ravanelli, Titouan Parcollet, Aku Rouhe, Peter Plantinga, Elena
  Rastorgueva, Loren Lugosch, Nauman Dawalatabad, Chou Ju-Chieh, Abdel Heba,
  Francois Grondin, William Aris, Chien-Feng Liao, Samuele Cornell, Sung-Lin
  Yeh, Hwidong Na, Yan Gao, Szu-Wei Fu, Cem Subakan, Renato De~Mori, and Yoshua
  Bengio,
\newblock ``Speechbrain,'' \url{https://github.com/speechbrain/speechbrain},
  2021.

\bibitem{richey2018voices}
Colleen Richey, Maria~A. Barrios, Zeb Armstrong, Chris Bartels, Horacio Franco,
  Martin Graciarena, Aaron Lawson, Mahesh~Kumar Nandwana, Allen Stauffer,
  Julien van Hout, Paul Gamble, Jeff Hetherly, Cory Stephenson, and Karl Ni,
\newblock ``Voices obscured in complex environmental settings (voices)
  corpus,''
\newblock {\em arXiv:1804.05053}, 2018.

\bibitem{fu_metricgan_2021}
Szu-Wei Fu, Cheng Yu, Tsun-An Hsieh, Peter Plantinga, Mirco Ravanelli, Xugang
  Lu, and Yu~Tsao,
\newblock ``{MetricGAN}+: An improved version of {MetricGAN} for speech
  enhancement,''
\newblock {\em {arXiv}:2104.03538}, 2021.

\bibitem{conformer}
Anmol Gulati, Chung-Cheng Chiu, James Qin, Jiahui Yu, Niki Parmar, Ruoming
  Pang, Shibo Wang, Wei Han, Yonghui Wu, Yu~Zhang, and Zhengdong Zhang,
\newblock ``Conformer: Convolution-augmented transformer for speech
  recognition,''
\newblock in {\em Interspeech}, 2020.

\bibitem{tsne}
L.J.P. {van der Maaten} and G.E. Hinton,
\newblock ``Visualizing high-dimensional data using t-sne,''
\newblock in {\em Journal of Machine Learning Research}, 2008, vol.~9, pp.
  2579--2605.

\bibitem{kingma2014adam}
Diederik~P Kingma and Jimmy Ba,
\newblock ``Adam: A method for stochastic optimization,''
\newblock in {\em ICLR}, 2015.

\bibitem{vaswani2017attention}
Ashish Vaswani, Noam Shazeer, Niki Parmar, Jakob Uszkoreit, Llion Jones,
  Aidan~N Gomez, {\L}ukasz Kaiser, and Illia Polosukhin,
\newblock ``Attention is all you need,''
\newblock in {\em NeurIPS}, 2017.

\bibitem{DesplanquesTD20}
Brecht Desplanques, Jenthe Thienpondt, and Kris Demuynck,
\newblock ``{ECAPA-TDNN:} emphasized channel attention, propagation and
  aggregation in {TDNN} based speaker verification,''
\newblock in {\em Interspeech 2020}, Helen Meng, Bo~Xu, and Thomas~Fang Zheng,
  Eds. 2020, pp. 3830--3834, {ISCA}.

\bibitem{Nagrani17}
Arsha Nagrani, Joon~Son Chung, and Andrew Zisserman,
\newblock ``Voxceleb: a large-scale speaker identification dataset,''
\newblock in {\em INTERSPEECH}, 2017.

\end{thebibliography}
\clearpage
\appendix
\section{Supplementary Materials}

In this supplementary material, we provide additional details about:
\begin{enumerate}[leftmargin=*,itemsep=0pt]
    \item Video (with audio) for demos of the collected data as well as  qualitative assessment of VIDA’s performance.
    \item Implementation details of our model and data pre-processing.
    \item Evaluation details of downstream tasks and corresponding metrics.
    \item Ablation on visual sensors.

\end{enumerate}

\subsection{Qualitative Video}
This video includes examples for audio-visual data in simulation and in the real-world. We demonstrate examples of our dereverbration model applied to these inputs. The video is available at \url{https://youtu.be/zPeAjlwo6XA}.

\subsection{Implementation Details} 
For the STFT calculation, we sample the input audio at 16 kHz and use a Hamming window of size 400 samples (25 milliseconds), a hop length of 160 samples (10 milliseconds), and a 512-point FFT. By retaining only the positive frequencies and segmenting the spectrograms into 256-frame chunks (corresponding to approximately 2.5 seconds of sound), the final audio input size to our UNet is 256x256. We use the Adam optimizer~\cite{kingma2014adam} to train our model with $lr=0.001$. We decay the learning rate exponentially to $lr=0.0001$ in 150 epochs. We set the batch size to 96 and train all models for 150 epochs, which is long enough to reach convergence. We set the margin $m$ to 0.5, phase loss weight $\lambda_1$ to 0.08 and matching loss weight $\lambda_2$ to 0.001.

\subsection{Evaluation Details}
We evaluate our model on three tasks: speech enhancement (SE), automatic speech recognition (ASR), and speaker verification (SV). 
\begin{itemize}
    \item For SE, the goal is to improve the overall sonic quality of the speech signal, which we measure automatically using the standard Perceptual Evaluation of Speech Quality (PESQ) ~\cite{rix_perceptual_2001} metric.
    \item For ASR, the goal is to automatically transcribe the sequence of words that were spoken in the audio recording. For this task, we report the Word Error Rate (WER), which is the standard metric used in ASR and reflects a word-level edit distance between a recognizer's output and the ground-truth transcription. 
    \item For SV, the goal is to detect whether or not two different spoken utterances were spoken by the same speaker. For SV, we report the Equal Error Rate (EER), a standard metric in the SV field indicating the point on the Detection Error Tradeoff (DET) curve where the false alarm and missed detection probabilities are equivalent. 
\end{itemize} Since the spectrogram MSE loss we optimize during training does not perfectly correlate with these three task-specific metrics, we perform model selection (across snapshots saved each training epoch) by computing the task-specific evaluation metric on 500 validation samples. We then select the best model snapshot independently for each downstream task and evaluate on the held-out test set; the same model selection procedure is also used for all of our baseline models.

For the ASR and SV tasks, we use the SpeechBrain~\cite{SB2021} toolkit. For ASR, we use the HuggingFace Transformer~\cite{vaswani2017attention} + Transformer LM model pre-trained on LibriSpeech~\cite{librispeech}. We evaluate this model off-the-shelf on our (de)reverberated version of the LibriSpeech test-clean set, and also explore fine-tuning the model on the (de)reverberated LibriSpeech train-clean-360 data.
For the SV task, we use SpeechBrain's ECAPA-TDNN embedding model~\cite{DesplanquesTD20}, pre-trained on VoxCeleb~\cite{Nagrani17}. For performing verification, we evaluate the model on a set of 80k randomly sampled utterance pairs from the test-clean set (40k same-speaker pairs, 40k different-speaker pairs) using the cosine similarity-based scoring pipeline from SpeechBrain's VoxCeleb recipe. In the verification task, we use the clean (non-reverberated) speech as the reference utterance, and the reverberant speech as the test utterance. As in the ASR task, we evaluate this model on our dereverberation model's outputs both off-the-shelf, as well as after fine-tuning on the (de)reverberated train-clean-360 set.

\subsection{Ablation on Visual Sensors}
To understand the importance of each input sensor, we ablate the RGB and depth input as shown in Table~\ref{tab:sensor_ablation}. Dropping either RGB or depth makes the WER worse.  We hypothesize that this is because they contain distinct information for the learning of room acoustics. The depth image is better for capturing room geometry, while the RGB image is better for capturing material and speaker location information.

In addition, we perform early fusion of RGB and Depth images by stacking them along the channel dimension (w/ early fusion in Table~\ref{tab:sensor_ablation}) and use one ResNet18~\cite{resnet18} model instead of two. This method also has worse WER, which validates our design choice of extracting RGB and depth features separately.

\begin{table}[t!]\hspace*{-0.2in}\small
\setlength{\tabcolsep}{4pt}
\centering
\begin{tabular}{ c| c | c | c}

 \toprule
                                                & \multicolumn{1}{c|}{\textit{SE}}
                                                & \multicolumn{1}{c|}{\textit{ASR}} & \multicolumn{1}{c}{\textit{SV}} \\
                                                & {PESQ $\uparrow$} & {WER (\%) $\downarrow$} & {EER (\%) $\downarrow$}\\
 \midrule
    Reverberant                 & 1.54 & 8.86  & 4.69 \\
    VIDA w/o VAN                & 2.32 & 4.92  & 4.67\\
    VIDA w/o RGB                & \bfseries 2.38 & 4.76 & \bfseries 3.82 \\
    VIDA w/o depth              & \bfseries 2.38 & 4.52  & 3.99 \\
    VIDA w/ early fusion        & \bfseries 2.38 & 4.56 & 3.94 \\
    VIDA                        & 2.37 & \bfseries 4.44  &  3.99 \\
 \bottomrule
\end{tabular}
\caption{Ablations on visual sensors. Percentages in parenthesis represent relative improvements over the reverberant baseline.
}
\vspace*{-0.05in}
\label{tab:sensor_ablation}
\end{table}



\end{document}